\documentclass[11pt]{article} 

\usepackage[margin=1in]{geometry} 
\usepackage{color}
\usepackage{soul}
\usepackage{indentfirst}
\usepackage{amsmath}

\usepackage{graphicx}

\usepackage{textcomp}

\usepackage{titlesec}
\titleformat{\section}{\Large\bfseries}{\thesection}{1em}{}[]
\titleformat{\subsection}[runin]{\bfseries}{\thesection}{}{}[]

\begin{document}

\begin{center}

{\Large\bf Microbuckling of Fibrin Provides a Mechanism for Cell Mechanosensing}
\vspace{22pt}

Jacob Notbohm$^{1*}$, Ayelet Lesman$^2$, Phoebus Rosakis$^{3}$, David A. Tirrell$^2$, Guruswami Ravichandran$^1$
\vspace{11pt}

$^1$Division of Engineering and Applied Science, California Institute of Technology, Pasadena, CA  91125 \\
$^2$Division of Chemistry and Chemical Engineering, California Institute of Technology, Pasadena, CA 91125 \\
$^3$Department of Theoretical and Applied Mathematics, University of Crete, Heraklion 70013, Greece

$^*$ Present address: Department of Engineering Physics, University of Wisconsin--Madison, Madison, WI 53706, jknotbohm@wisc.edu

\end{center}

\section*{Abstract}
Biological cells sense and respond to mechanical forces, but how such a mechanosensing proccess takes place in a nonlinear inhomogeneous fibrous matrix remains unknown. We show that cells in a fibrous matrix induce deformation fields that propagate over a longer range than predicted by linear elasticity. Synthetic, linear elastic hydrogels  used in many mechanotransduction studies fail to capture this effect. We develop a nonlinear  microstructural finite element model for a fiber network to simulate  localized deformations induced by  cells. The model captures  measured cell-induced matrix displacements from experiments and identifies an important  mechanism for long range cell mechanosensing: loss of compression stiffness due to microbuckling of individual fibers. We show evidence that cells sense each other through the formation of localized intercellular bands of tensile deformations caused by this mechanism. 

\section*{Key Words}
\noindent Fibrous matrix, Buckling, 3D traction force, Cell mechanics

\newpage

\section*{Introduction} 
Physical cues control cell behavior through various mechanisms collectively referred to as mechanotransduction \cite{1vogel}. For example, the stiffness of a cell's environment controls cellular morphology, migration, and development \cite{2discher}. Equally important is the response of cells to direct physical forces either through cell-cell adhesions \cite{3trepat,4notbohm} or through the extracellular matrix \cite{5lo,6reinhart,7tang}. Nearly all previous work on mechanotransduction has used synthetic, linear elastic gels \cite{8kandow}. The mechanical properties of physiological extracellular environment, however, deviate entirely from simple homogeneous linear elasticity. Natural fibrous matrices exhibit strain stiffening \cite{10storm}, tensile normal strains under shear loading \cite{11janmey}, negative compressibility \cite{12brown}, and lower stiffness in compression than in tension \cite{11janmey}. These nonlinear properties of biological gels can have a dramatic effect on behaviors like cell spreading \cite{9winer}.

Various models have simulated nonlinearity of fibrous biological materials, but relatively few have considered local, non-uniform deformations in such nonlinear inhomogeneous materials \cite{16shokef,notbohmthesis}. Instead, nearly all previous studies have focused on homogeneous shearing  \cite{10storm,13conti,14onck,15heussinger} or uniaxial tension \cite{12brown} of the bulk material. These studies of uniform deformations have revealed novel constitutive behavior of fibrous materials, but they fail to simulate deformations similar to those applied by a cell. By contracting and changing shape, cells apply localized forces to their surroundings, resulting in inhomogeneous stress and deformation fields in the matrix. Given the lack of theoretical and experimental studies of cell--matrix interactions at the local scale, there remains a need to quantify cell-generated forces and displacements and to discern how cells respond to nonlinear properties of fibrous materials at the scale sensed by the cell.

Here we experimentally measure 3D cell-induced matrix displacements and report two findings: (i) displacements decay much slower with distance from the cell than predicted by linear elasticity; (ii) multiple cells cause localized  matrix densification and fiber alignment in tether-like bands joining them. We hypothesize that the mechanism responsible for these phenomena is loss of compression strength due to microbuckling of individual fibers. To test this claim, we develop a microstructural finite element (FE) network model of the fibrin matrix. Buckling of individual fibers is modeled by a loss of  stiffness in compression for network elements. Our model agrees with previous experimental observations for fibrin, and it predicts both the slow decay of displacements and localization of intercellular tethers. Variants of the model without loss of stiffness in compression fail to predict these effects. The long range of cell-generated displacements and stresses, and the localization into intercellular tensile tethers, allow cells to sense each other and their surroundings over larger distances through a fibrous matrix than through homogeneous hydrogels with linear elastic behavior. We show evidence that cells respond to localized tension by growing protrusions towards one another, guided by the dense aligned fibers in tethers. This points to fiber microbuckling as an important mechanism responsible for enhanced range of cell mechanosensing in fibrous matrix environments.

\section*{Results}

\subsection*{Cell-Induced Matrix Displacements.}
We motivate our model by first considering cell-induced displacements within a 3D fibrous matrix during initial cell spreading. A cell seeded in a 3D matrix initially applies tensile tractions to the fibers by undergoing uniform isotropic contraction while in an essentially  spherical state. This suggests Eshelby's solution for a contracting spherical inclusion in a homogeneous, linear elastic, infinite medium \cite{17eshelby}, as a simple analytical model for cell-induced matrix deformation.  In this  solution, the  displacement  magnitude $u=u(r)$ scales as $u(r) \sim r^{-2}$  with distance $r$ from the cell center. Stress components, e.g., the radial component  $\sigma_{rr}$, scale as $\sigma_{rr}\sim r^{-3}$. A spreading, elongated, ellipsoidal cell with polarized alignment, applies tractions equivalent to equal and opposite forces at its poles, i.e. a dipole \cite{18schwarz}, in view of force equilibrium. Displacements due to  a dipole in a 3D linear elastic continuum also scale as $u \sim r^{-2}$. One would thus expect  displacements induced by a  spreading cell in a 3D matrix to scale similarly.
  
Using confocal microscopy and digital volume correlation \cite{notbohmthesis,26franck,lesmannotbohm}, we measure displacements induced by isolated fibroblast cells embedded in a 3D fibrin matrix. Displacements induced by the cells are largest near to the cell and decrease with distance from the cell (Fig.~1a). We quantify the rate at which displacements decay over distance by computing displacements along linear paths starting at the the center of cell and ending $\sim$100 \textmu m away (Fig.~1a, white line). Experimental data from multiple different cells are plotted on a logarithmic scale in Fig.~1b. Data are fit to the form $u(r)=Ar^{-n}$. Here $A$  and $n$ are constants; $n>0$ is a decay power. The larger the value of $n$, the faster the displacement $u$ decays  with distance $r$ from the cell center. Fits of the experimental data yield $n=0.52$ (mean over data from 6 cells during multiple time points), indicating that displacements decay much slower than predicted by the linear elastic solution $n=2$. 
The ratio of the RMS errors of fits to $u \sim r^{-0.52}$ and $u \sim r^{-2}$, is $0.14\pm ± 0.07$ (mean $\pm$ standard deviation), hence the scaling $u \sim r^{-0.52}$ describes cell-induced displacements in a fibrin matrix far better than the 3D linear elastic scaling $u \sim r^{-2}$.

A striking difference between fibrin networks and homogeneous gel matrices is 
the phenomenon of buckling of individual fibers under compression; an example is shown in Fig.~5 of ref. \cite{kim}. This is directly responsible for the decreased ability of  fibrin networks to sustain compressive stresses. Each fiber has very low resistance to bending, much like a flexible string \cite{23piechocka}. If one pulls at the ends of a string, it resists tension. If one pushes the ends of a string towards each other, the string bends easily without resisting compression (i.e., it buckles), and this buckling can change the mechanical response of a network \cite{bischofs2008BiophysJ}. Fibrin exhibits a larger stiffness in tension than compression \cite{11janmey,21stylianopoulos} due to buckling of individual fibers  under compression \cite{13conti,kim,19vader,20munster}. Is this nonlinearity responsible for the discord between the observed displacement scaling and  the prediction based on a linear elastic matrix assumption?  While it may be possible to address this via a continuum model  for a material with lower stiffness in compression \cite{rosakis}, here we present a simple theoretical continuum argument, which we will investigate in detail using a discrete model. Since the cell exerts radial contractile traction forces, the stress tensor in the matrix has a tensile (positive) radial component in 3D spherical coordinates, and two contractile (negative) hoop (angular) components. Assuming the individual fibers of the fibrin matrix buckle under a small compressive load, the contractile hoop components of the stress tensor are small and can be neglected. This assumption reduces the radial equilibrium equation \cite{love}  to
\begin{equation}
\frac{d\sigma_{rr}}{dr} +2\frac{\sigma_{rr}}{r} =0.
\end{equation}
Solving Eq.~(1) gives $\sigma_{rr} \sim r^{-2}$. Thus, stress due to cell contraction is transmitted over a longer range than under the scaling $\sigma_{rr} \sim r^{-3}$ predicted by linear elasticity. Assuming piecewise linear stress--strain relations with zero stiffness in compression, $\sigma_{rr} $ is proportional to the radial strain $ du(r)/dr$   which gives $ u \sim r^{-1}$. (Coupling between $\sigma_{rr}$ and the hoop strains $\gamma_{\theta\theta}$, $\gamma_{\phi\phi}$ vanishes due to  hyperelastic reciprocity: $\partial \sigma_{rr}/\partial\gamma_{\theta\theta}=\partial \sigma_{\theta\theta}/\partial\gamma_{rr}=0$  since $ \sigma_{\theta\theta}=0$ in the compressive regime. For more details on a hyperelastic material model that leads to Eq.~(1) as a special case, see ref. \cite{rosakis}.)  The scaling from this simple analysis, $u \sim r^{-1}$, points towards displacements that propagate over a longer range than the 3D linear elastic scaling, $u \sim r^{-2}$. Furthermore, the scaling from the theoretical analysis is closer to the experimentally observed scaling, $u \sim r^{-0.52}$ than to the linear elastic one. This plausibility argument shows the right trend, but ignores the inhomogeneous and discrete nature of the fibrin network. To account for these factors, we turn to a microstructural network model.

\subsection*{Model.} We develop a FE-based microstructural model consisting of a 2D/3D network of linear elements representing fibers. This model expands on one that we have recently developed \cite{notbohmthesis}. Each element undergoes uniaxial tension/compression and rotates with no resistance. We model buckling of fibers  as a loss of stiffness in compression in the stress--strain relation of individual elements. This agrees qualitatively with observed behavior in similar systems \cite{lakes}. In the context of our model, ``microbuckling'' will refer to elements obeying a stress--strain relation where the stiffness (slope) under compression is smaller than the stiffness under tension; see Fig.~2a, blue line. In the following simulations, we use a ratio of stiffness in compression to stiffness in tension $\rho=0.1$. While the choice of $\rho=0.1$ is arbitrary, we find that any positive ratio of stiffnesses $\rho$ significantly less than unity yields similar results. In contrast, ``no microbuckling'' will refer to elements with $\rho=1$, i.e. elements with a linear stress--strain relation without a reduced compression stiffness. For most simulations, networks comprise  elements with a bi-linear stress-strain relationship (Fig.~2a, different slopes in tension and compression). We will also account for the possibility of entropic elasticity by employing a wormlike chain-type (WLC) stress-strain relationship \cite{10storm, 23piechocka}, where the stiffness increases with strain in tension (Fig.~2b). The elements connect an array of nodes as in Fig.~2c. Randomness is added to  nodal positions to simulate the random array of fibers of different lengths typical of a fibrous network (Fig.~2d).

Another important aspect of actual fibrin networks is their low connectivity, or coordination number $C$, i.e. the average number of fibers meeting at a node. The network of Fig.~2c,d has  $C=8$, while actual fibrin often has a typical value of $C=3$ \cite{wyart}. This is below the critical value for rigidity, $C=6$ or $4$ for 3D and 2D networks, respectively. As a result, fibrin is typically a ``floppy'' network, and this affects its mechanical properties \cite{wyart}. To obtain a model network with lower connectivity  (such as $C=3$ in Fig.~2e), we removed elements at random from the original $C=8$ network of Fig.~2d.  As in ref. \cite{wyart}, deleted elements were replaced by weak elements, whose stiffness was six orders of magnitude less than that of the deleted ones; this ensured stability of numerical calculations.

In contrast to previous models that focus on the  macroscale behavior of a fiber network \cite{13conti,14onck,15heussinger}, we simulate the inhomogeneous, localized displacements induced in a fibrin matrix by an embedded cell. We begin with 2D FE simulations where the cell is modeled as a contracting circle. The matrix occupies the region $a<r<b$, where $r$ is distance from the cell center; here $a$ is the cell radius, and $b/a=50$. The outside boundary $r=b$ is free (a zero traction boundary condition is imposed). The cell boundary $r=a$ undergoes a radial contractile displacement $u(a)=-0.1a$. Simulations were performed for different connectivities in the interval $2.5\le C\le 8$ for bilinear element networks with microbuckling and without. The displacement magnitude $u$ was computed (Fig.~3a), averaged around the circular region (Fig.~3b, Supplemental Fig.~S1a), and fit against distance from the center of the circular region $r$ to $u=Ar^{-n}$ for the constants $A$ and $n$. Results, $n$ plotted versus connectivity $C$, are shown in Fig.~3c. In general, the decay power $n$ for networks with microbuckling (Fig.~3c, open circles) is  substantially lower, by at least $0.4$,  than for networks without microbuckling of fibers (Fig.~3c, black circles). This is true for a wide range of connectivities, with an exception near the critical value $C=4$; for these values $n\approx 0.6$ in both types of networks. We observe larger spatial inhomogeneities of displacement at the scale of individual fibers in networks with $C=4$ than in those with both subcritical and supercritical connectivity (Fig.~3b, Supplemental Fig.~S2). These fluctuations are due to the change in phase from subcritical to supercritical connectivity as detailed elsewhere \cite{wyart,broedersz2011}. For the case without microbuckling (i.e. with linear stress--strain relation) since individual elements have linear stress--strain behavior, we compare the displacement to the linear elastic 2D solution $u=Ar^{-n}+Br^{n}$ for the constants $A$, $B$, and $n$. Except near $C=4$, we find $n=0.89\pm 0.04$ (mean $\pm$ standard deviation, essentially independent of $C$ over all connectivities). This value of $n=0.89$ is close to the 2D linear elastic solution $n=1$. Connectivity does not appear to play a major role in displacement decay except near the critical value. We find no change in these conclusions when the zero traction boundary condition is replaced by a zero displacement condition fixing the external boundary; see Supplemental Fig.~S3. Thus we conclude microbuckling is crucial for the slow decay of displacements.

The long range of cell-induced displacements has been previously attributed to strain stiffening \cite{9winer}, but this has been disputed \cite{25rudnicki}. We observe that the experiments of ref. \cite{9winer} were performed on fibrin, which exhibits microbuckling. Also, ref. \cite{25rudnicki} provides evidence against strain stiffening as the underlying mechanism, but does not seem to propose an alternative. To help settle this, we repeated our simulations with elements whose stress--strain curve is of WLC type and stiffens in tension (Fig.~2b). Two versions of stiffening WLC stress--strain curves were compared. A curve whose slope is continuous at zero strain and increases smoothly in tension models a tension-stiffening material that does not undergo microbuckling (black dashed line in Fig.~2b). The alternative stress--strain curve has a discontinuous slope at zero strain (10 times smaller than the tangent stiffness for small tensile strain). It models microbuckling  (red solid line in Fig.~2b) combined with tension stiffening. In all cases, values of the decay exponent $n$ from fits for WLC networks (Fig.~3d,e,f, Supplemental Fig.~S1c) agreed well with fits for bilinear networks of the same connectivity and same (buckling or non-buckling) type (Fig.~3c). This provides strong evidence that the tension-stiffening nonlinearity in the absence of microbuckling is not the cause of the slow displacement decay that we observe.

Until now, we have considered round cell geometries, which do not capture the elongated shape of spread cells. For an anisotropic cell contracting along its long axis, an ellipsoid more accurately captures the cell's shape. For this geometry, linear elasticity predicts that displacements far from the cell scale as $u_1 \sim x_1^{-n}$ where $n=2$ in three dimensions, $n=1$ in two dimensions; $x_1$ is the distance along the major axis from the center of the ellipsoid (or ellipse in two dimensions), and $u_1$ is the displacement in the $x_1$ direction \cite{17eshelby}. To compare with the linear elastic solution, we placed in our fibrous network model an ellipse with a ratio of semi-major and semi-minor axes $a_1/a_2=4$. As with the contracting circle, the matrix occupied a circular region of radius $b=50a$ with the nodes on the boundary $r=b$ free and $a$ defined for the ellipse as $a \equiv \sqrt{a_1 a_2}$. Contractile displacements  were applied on the boundary of the ellipse, with nonzero component  $u_1(x_1)= - 0.1a (x_1/a_1)$  (along the long axis of the ellipse). This is equivalent to subjecting the ellipse to a negative uniaxial strain  along the ellipse's long axis. The largest magnitude of contractile displacement is $|u_1(a_1)| = 0.1a$ (at the ellipse tip), the same value as for the contracting circle. Displacements along the axis of the ellipse (Fig.~4a) appear to scale similarly to the displacements induced by the contracting circle (Fig.~3b). Indeed, the fittings to $u_1=Ax_1^{-n}$ show decay powers $n$ that are significantly smaller for networks with microbuckling (Fig.~4b, $\rho=0.1$) than without (Fig.~4b, $\rho=1$, Supplemental Fig.~S1b). Like the contracting circle, the ellipse exhibits an exception at the critical connectivity $C=4$. The trend shown in Figs. 3 and 4 is clear: microbuckling results in cell-induced displacements that propagate over a longer range than predicted by linear elasticity for both a contracting circle and a contractile ellipse.
 
We also performed 3D simulations (contracting spherical cell), with similar conclusions. We recall that the 3D linear elastic solution predicts $u\sim r^{-2}$. The theoretical argument based on Eq.~(1) gives $u\sim r^{-1}$, while a fit to our experiments yields $u\sim r^{-0.52}$.  For 3D networks (microbuckling bilinear elements) with $C=14$, a fit to $u=Ar^{-n}$ gives $n=0.67$. For $C=3.5$ (below the critical value for rigidity  $C=6$) we found $n=0.82$ (Supplemental Fig.~S4). These results combine to show that microbuckling of fibers is the key mechanism responsible for the longer range of cell-induced deformations in a fibrin matrix.

\subsection*{Tethers.} 

Can cells exploit the long  propagation range of matrix deformations they themselves induce for sensing the presence of other cells?  We use confocal microscopy to visualize both the matrix and multiple fibroblast cells embedded in it. We observe that cells whose distance from each other is of the order of 10 cell diameters are connected to each other by linear bands consisting of aligned and densely packed matrix fibers (Fig.~5). Within these ``tethers'' fibers appear to be in tension in the direction joining the cells. These tethers also occur between multicellular explants in a fibrous matrix \cite{stopak1982,sawhney2002}, but the mechanism for their formation remains unknown. The tethers extend far beyond a single cell's protrusion (Fig.~5). Matrix remodeling by degradation or deposition cannot be responsible for alignment at such a large distance from the cell. This leads us to examine the hypothesis that tethers form due to tensile forces.

A previous model has shown that a point force in a fibrous medium induces forces which propagate through tether-like paths \cite{HeussingerFrey2007}. The point force loading of this previous model was not intended to simulate forces due to cells, which maintain force equilibrium while pulling on the matrix. To investigate the physical mechanism of tether formation, we used our FE network model with microbuckling to simulate a pair of contracting cells. A symmetric boundary condition is imposed at the bottom of a square region containing a circle of radius $a$ (the other boundaries are free of applied tractions). By symmetry this is equivalent to a pair of identical contracting circular cells in the fibrous matrix. We apply an isotropic inward radial displacement of $0.1a$ to the circular region. For a cell with a radius of $a=10$ \textmu m, this value of $0.1a$ corresponds to 1 \textmu m, in agreement with the experimental data (Fig.~1b).

A different model \cite{shenoy} requires cell displacements nearly an order of magnitude higher than the experimentally observed value of 1 \textmu m in order to predict appreciable interaction between cells. The simulated tensile strains in our network occur almost entirely in the band between the two cells, along aligned linear paths formed by elements in tension (Fig.~6a). Compressive strains localize perpendicular to these tensile tethers (Fig.~6b). Due to low compression stiffness (microbuckling), the magnitude of compressive strain in elements roughly perpendicular to the tether is more than twice the magnitude of the tensile strain. Thus within the tether, the trace of the strain tensor, or the volumetric strain, is negative, consistent with the observation that matrix fiber density increases between pairs of cells (Fig.~5). When simulating networks without microbuckling, we found no such tethers forming; instead, tensile strains had a nearly radially symmetric distribution around each cell (Supplemental Fig.~S5). Thus we conclude that localization of matrix deformation caused by multiple cells in the tensile, tether-like regions joining those cells occurs because of microbuckling of fibers normal to the tethers.

\section*{Discussion}

We have shown that cells embedded within a fibrin matrix exert forces that cause matrix displacements to propagate over a longer range than predicted by linear elasticity. The long range propagation of displacements has been previously observed for cells on a flat, 2D fibrous substrate \cite{9winer} and for multicellular constructs in a 3D matrix \cite{gjorevski2012}. Our observations, first reported in ref. \cite{notbohmthesis}, confirm this result for single cells in a 3D system. Here, we further quantify the spatial decay of displacements by fitting to a power law resulting in displacements scaling as $u \sim r^{-0.52}$. While the propagation of displacements over a long range is now apparent, the precise mechanism is still unclear. Recent studies have argued for \cite{9winer} and against \cite{25rudnicki} the hypothesis that long range propagation of displacements results from strain stiffening. When we included strain stiffening in the behavior of fibers, but suppressed compression weakening due to buckling, long range propagation was not observed in simulations of our model. Thus we conclude that fiber buckling---rather than strain stiffening in tension---explains the long range propagation of displacements observed in the experiments.

To simulate buckling, we used a model that does not resist changes in angle between the elements. Previous work \cite{broedersz2011} has pointed to bending as an important mechanism that controls the mechanical response of fibrous materials. However, the model of ref. \cite{broedersz2011} does not allow for buckling or even bending of individual fibers. Instead it models bending by penalizing changes in angle between initially co-linear elements that meet at a node. Moreover, even for models that include bending, forces are dominated not by transverse bending displacements but by axial ones \cite{HeussingerFrey2007}. To address here the question of microbuckling, we focus on these axial displacements. We assume fibers buckle immediately after a compressive load is applied, i.e., we assume the fiber buckling load (equivalently the buckling strain) is equal to zero. Is this assumption reasonable? A typical fibrin fiber with a length of 1 \textmu m, diameter of 0.2 \textmu m, persistence length of 40 \textmu m \cite{23piechocka}, and Young's modulus of 15 MPa \cite{collet2005} will buckle at a compressive strain of $\sim\!\!4\times10^{-4}$\%. This value is small compared to typical strains experienced in the matrix ($\sim\!\!1$\%), so our choice of setting the transition point between different stiffnesses at the onset of compression  (vanishing buckling load) is justified.

Besides long range propagation of displacements, fibrous materials exhibit what is termed in ref. \cite{12brown} a ``negative compressibility'' in uniaxial tension, i.e., a negative ratio of the trace of the stress tensor and the trace of the strain tensor during a uniaxial tension experiment.
Fibrous materials also exhibit tensile normal stresses under prescribed  shear deformation \cite{11janmey}. This is in essence equivalent to negative normal (compressive) strains  when the material is subjected to external tangential forces (prescribed shear stress), but not constrained to expand or contract. To test whether our fiber model is consistent with these experimental observations, we simulated homogeneous uniaxial tension. We found that when fibers buckle, the model exhibits negative compressibility in tension.  In addition, under applied tangential forces equivalent to an external shear stress, the model responded with negative normal strains in shear. When microbuckling is removed from the model, neither of the aforementioned behaviors occurs (Supplemental Fig.~S6). Thus, our model with fiber microbuckling is consistent with previous experimental work on fibrin \cite{11janmey,12brown} and collagen \cite{19vader}. Certainly fibrous materials exhibit nonlinear behaviors besides microbuckling in compression, but our model points to microbuckling as being both consistent with previous experimental work and of major importance to the mechanical response of fibrous materials.

Together, our simulations and experiments reveal that microbuckling of fibrin enables cells to induce displacements that follow linear, tether-like paths that lead to other cells. These displacements propagate over a dramatically longer range than in a linear material. A remaining question of biological relevance is whether cells physically respond to the formation of tethers. In our experiments, we have observed pairs of cells forming pointed protrusions along these tethers and subsequently growing toward one another by several cell diameters (Fig.~5), sometimes eventually joining two cells (Supplemental Fig.~S7). A different model indicated that elongated cells initially pointed toward one another may sense displacements induced by their neighbors \cite{shenoy}, but it did not answer the question of how cells break their initial spherical symmetry to spread toward one another as observed in our experiments. Our model, which we present here and have described previously \cite{notbohmthesis}, suggests a mechanism whereby cells can sense one another during the initially spherical state. Even if each cell is initially spherical and contracts isotropically, the tether formation mechanism that we describe results in greater tension and fiber density that is highly polarized in the  direction of neighboring cells (Fig.~5). Both tension and fiber density may provide a directional signal: by growing protrusions along the direction of the tethers, cells have a higher chance of approaching one another. The fact that cells change shape and grow along such tethers supports the hypothesis that they use this very same mechanism to sense and even approach their neighbors. We expect future work to further clarify how cells sense the mechanical properties of fibrous materials and how we can better design artificial cell culture platforms to better control cellular response to forces within the extracellular matrix.

\section*{Methods}

\subsection*{Cell culture and matrix preparation.} 3T3 fibroblast cells stably expressing a green fluorescent protein--actin fusion protein were cultured in Dulbecco's Modified Eagle Medium supplemented with 10\% fetal bovine serum and 1$\times$ non-essential amino acids. Fibrin was fluorescently labeled by mixing fibrinogen (Omrix Biopharmaceuticals, Israel) and 546 Alexa Fluor (Life Technologies, Carlsbad, CA, USA) for 1 hour before filtering with a HiTrap desalting column (GE Healthcare, Milwaukee, WI, USA). Cell--fibrin constructs were created by suspending the cells in 20 U/mL thrombin solution (Omrix), mixing with 5 mg/mL labeled fibrinogen solution, and placing on a \#1.5 coverslip.

\subsection*{Microscopy and cell-induced matrix displacements.} Within 1 hour of seeding, cell-matrix constructs were transferred to a custom built 5\% CO$_\mathrm{2}$, 37$^{\circ}$C microscope enclosure. Imaging was performed with a Swept Field confocal microscope using a 40$\times$ NA 1.15 water immersion objective (Nikon Instruments, Melville, NY, USA). Volume stacks of the cells and fibrin matrix were captured every 15 minutes over time periods of several hours.

3D matrix displacements were computed directly from the images of the labeled fibrin using digital volume correlation \cite{26franck} with the initial volume stack (before cell spreading) taken as a reference for the correlation. The digital volume correlation software, written in Matlab (The Mathworks), is freely available online \cite{franckdvc2}. Propagation of cell-induced matrix displacements was quantified by computing displacement magnitudes along multiple linear paths propagating outward from the center of each initially rounded cell. To reduce errors caused by inhomogeneities within the matrix, displacements were averaged over $\sim\!\!3$ different paths and over $\sim\!\!10$ time points for each cell. After averaging, the standard deviation of the noise level was found to be 0.04 \textmu m.

\subsection*{Microstructural model.} The microstructural model was developed in the FE software Abaqus 6.10 (Dassault Systemes, Waltham, MA). Rod elements supporting tension and compression but not bending were connected as shown in Fig.~2c. Elements were randomly deleted to reduce the network connectivity. Removed elements were replaced by elements with stiffness six orders of magnitude smaller than the original elements. The choice of using weak elements with stiffness six orders of magnitude smaller than the original elements came after a series of convergence studies showed that further reduction in the stiffness of the weak elements had no effect on the displacements. Under tension, both a linear (Fig.~2a) and a strain-stiffening WLC relationship were investigated (Fig.~2b). Under compression a linear stress--strain relationship was used with slope given by $\rho$ times the slope at small tensile strains with $\rho=0.1$ for microbuckling and $\rho=1$ for no buckling. Strains within each element (as plotted in Fig.~5, Supplemental Fig.~S5) were computed by taking the natural logarithm of the stretch ratio, defined as the final element length divided by the initial element length. The 3D model used element connectivity as shown in Supplemental Fig.~S4 and the bilinear stress--strain relationship. Uniaxial tension was simulated in a square region by applying displacements on the top side, a symmetric boundary on the bottom side, and traction free boundaries on the right and left sides. Shear loading was simulated by applying horizontal displacements to the top of a thin rectangular region (aspect ratio 1/10) with a fixed bottom boundary and traction free conditions on the right an left. Apparent strains were computed by numerically computing the displacement gradients using a linear fitting. Effective Poisson's ratio was defined as the opposite of the ratio of apparent strains in the transverse and axial directions. 

\section*{Competing Interests Statement}
\noindent We have no competing interests.

\section*{Author Contributions}
\noindent J.N. and A.L. performed the experiments. J.N. performed the simulations. J.N. and P.R. wrote the manuscript. All authors discussed the results and gave approval for publication.

\section*{Funding Statement}
\noindent This work was funded in part by a grant from the National Science Foundation (Division of Materials Research No. 0520565) through the Center for the Science and Engineering of Materials at the California Institute of Technology, and in part by National Science Foundation Grant No. DMR-1206121. J.N. was supported by the National Science Foundation Graduate Research Fellowship under Grant No. DGE-1144469.


\bibliographystyle{unsrt}
\bibliography{notbohm_refs}


\newpage
\section*{Figures}
\vspace{11pt}

\begin{figure}[h!]
\begin{center}
\includegraphics[width=3in,keepaspectratio=true]{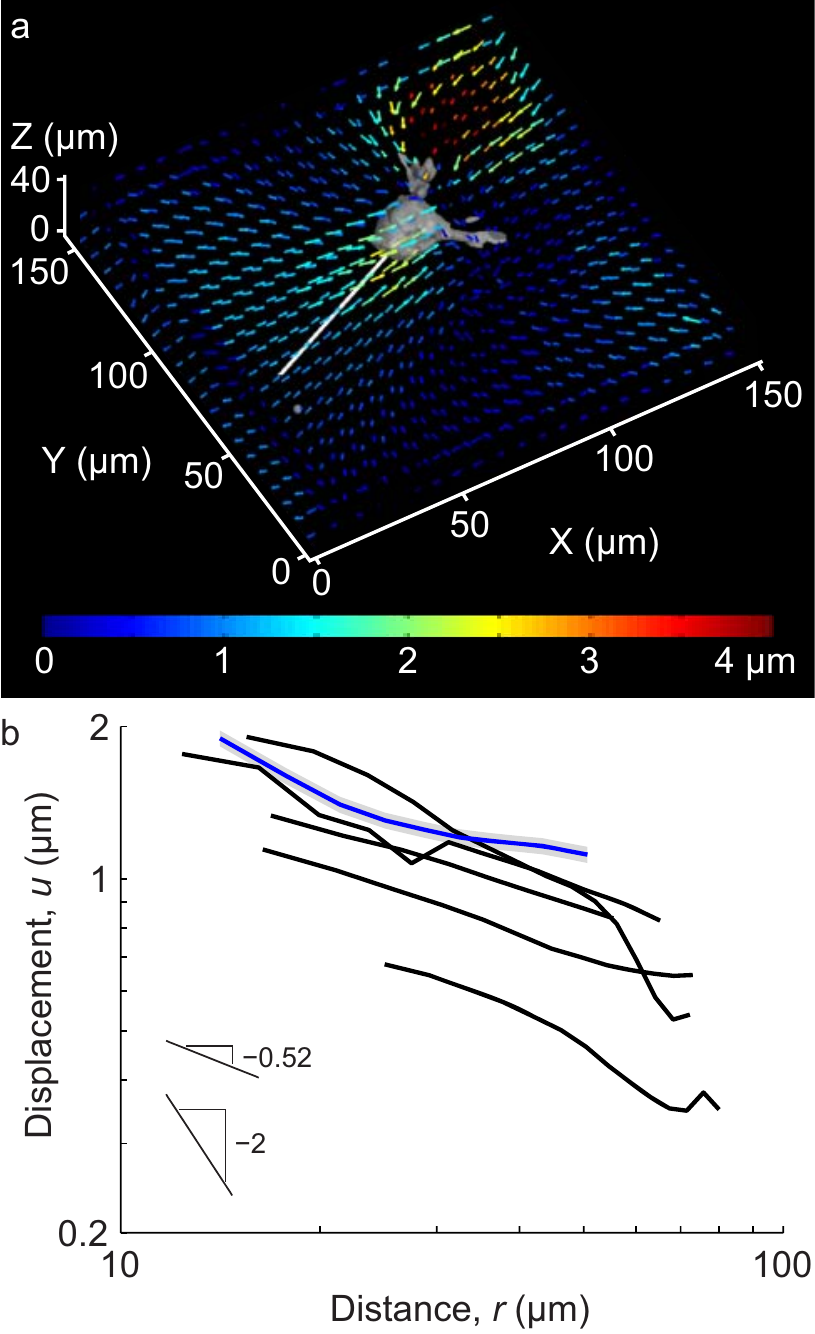}
\end{center}
\end{figure}
\noindent\textbf{Figure 1}: Experimentally measured displacements induced by isolated cells embedded within a 3D fibrous matrix. (a) The colored quivers plot 3D matrix displacement vectors applied by a cell to a 3D fibrin matrix. Paths (white) are chosen proceeding outward from the cell body. (b) Displacement magnitudes along the paths are averaged for multiple time points and plotted. Each curve is for a different cell. The blue curve shows displacements for the cell in (a). The gray shading behind the blue curve shows typical error of the displacement measurement after averaging. Data used to generate these curves is in the Supplemental Data.

\newpage
\begin{figure}[h!]
\begin{center}
\includegraphics[width=4in,keepaspectratio=true]{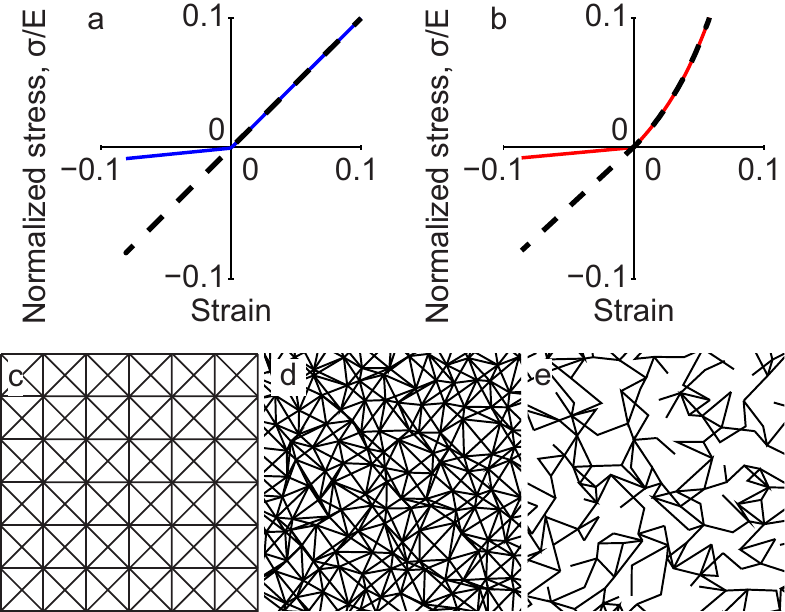}
\end{center}
\end{figure}
\noindent\textbf{Figure 2}: Finite element model network details. (a) Stress--strain curves for bilinear model. Stress $\sigma$ is normalized by Young's modulus $E$. Dashed black: linear without microbuckling ($\rho=1$); solid blue: bilinear with microbuckling ($\rho=0.1$). (b) Normalized stress--strain curves for the strain-stiffening model, which exhibits WLC-like behavior in tension. For this model $\rho$ is defined as the slope upon approaching the origin from the left divided by the slope upon approaching the origin from the right. A continuous slope at the origin ($\rho=1$) was used to simulate non-buckling elements (dashed black line) and a discontinuous slope ($\rho=0.1$) was used to simulate microbuckling elements (solid red line). (c) Network array. (d) Randomized network, $C=8$. (e) Network with reduced connectivity, $C=3$.

\newpage
\begin{figure}[h!]
\begin{center}
\includegraphics[width=4.35in,keepaspectratio=true]{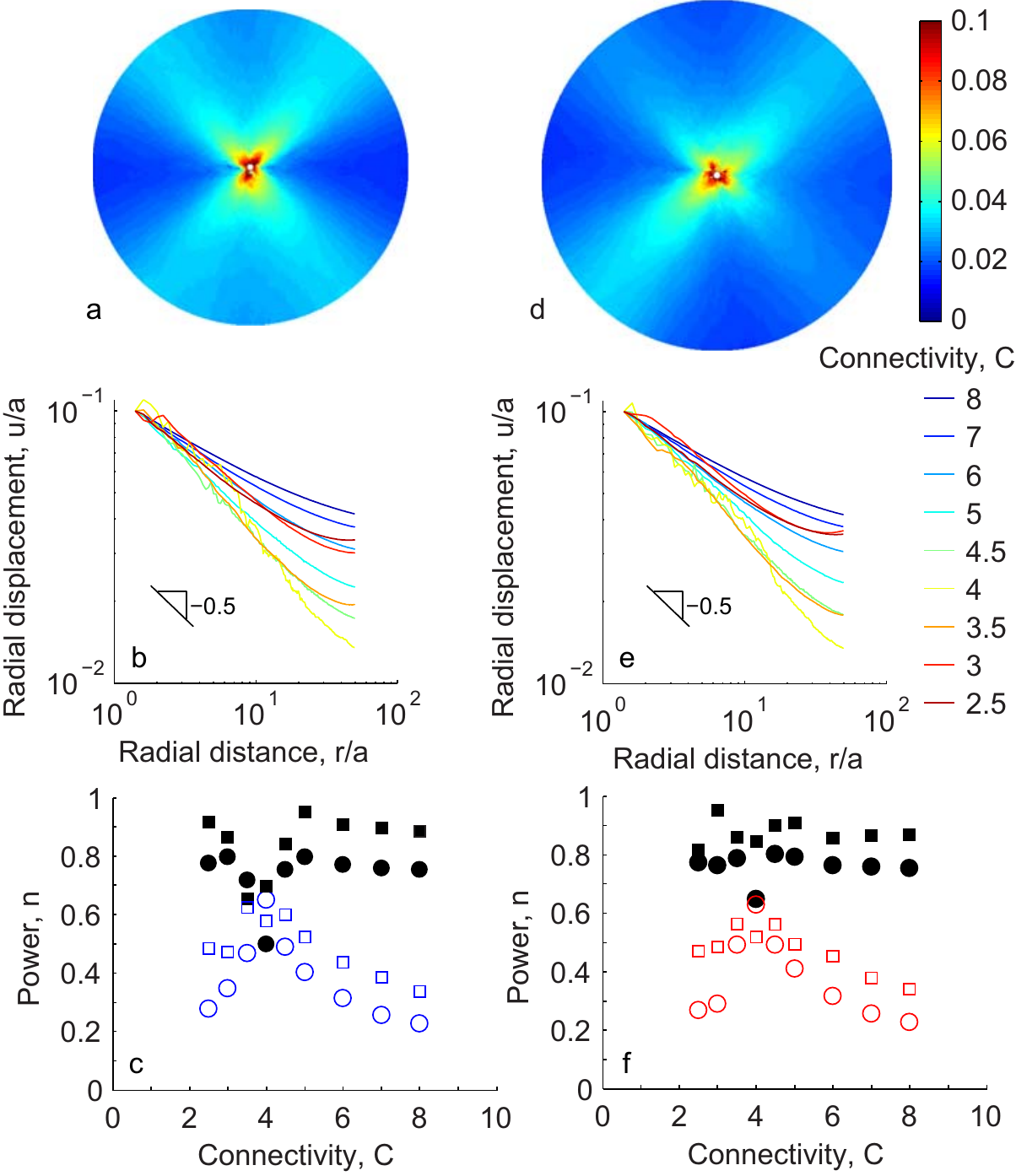}
\end{center}
\end{figure}
\noindent\textbf{Figure 3}: Long range propagation of displacements is due to microbuckling. (a) Inhomogeneous forces, like those applied by a cell, are modeled by a circle of radius $a$ contracting in a circular region with radius $b=50a$. Contours of normalized displacement $u/a$ are shown here for the case of the bilinear model (Fig.~2a) with microbuckling ($\rho=0.1$) and $C=3$. For a cell of radius 10 \textmu m, the applied displacement $u/a=0.1$ would correspond to 1 \textmu m. (b) Displacements are averaged around a circle of radius $r$ about the center of the model and plotted for simulations that used different connectivities ranging from $C=2.5$ to $C=8$. All curves show long range propagation of displacements with slopes of $\approx\!\!-0.5$. At the critical connectivity, $C=4$, displacements exhibit spatial inhomogeneities, resulting in fluctuations. (c) Decay power $n$ vs. connectivity $C$. Circles show fits to $u=Ar^{-n}$; squares show fits to $u=Ar^{-n}+Br^n$. Solid black symbols represent fibers that do not buckle ($\rho=1$); open symbols represent fibers that do buckle ($\rho=0.1$). Most powers $n$ for the case of microbuckling $\rho=0.1$ are $\approx\!\!0.5$, in agreement with the slope of $-0.5$ observed in panel (b). The value of $n \approx 0.5$ indicates displacement propagate over a longer range than predicted by linear elasticity, for which $n=1$ in two dimensions. Simulations are repeated for the strain stiffening WLC-type relationship (Fig.~2b). (d) Contours of displacement $u/a$ for the strain stiffening relationship with microbuckling ($\rho=0.1$). (e) Averaged displacements and (f) decay powers $n$ for the strain stiffening relationship. As in (c), circles show fits to $u=Ar^{-n}$; squares show fits to $u=Ar^{-n}+Br^n$. Solid black symbols represent fibers that do not buckle ($\rho=1$); open symbols represent fibers that do buckle ($\rho=0.1$).

\newpage
\begin{figure}[h!]
\begin{center}
\includegraphics[width=2.5in,keepaspectratio=true]{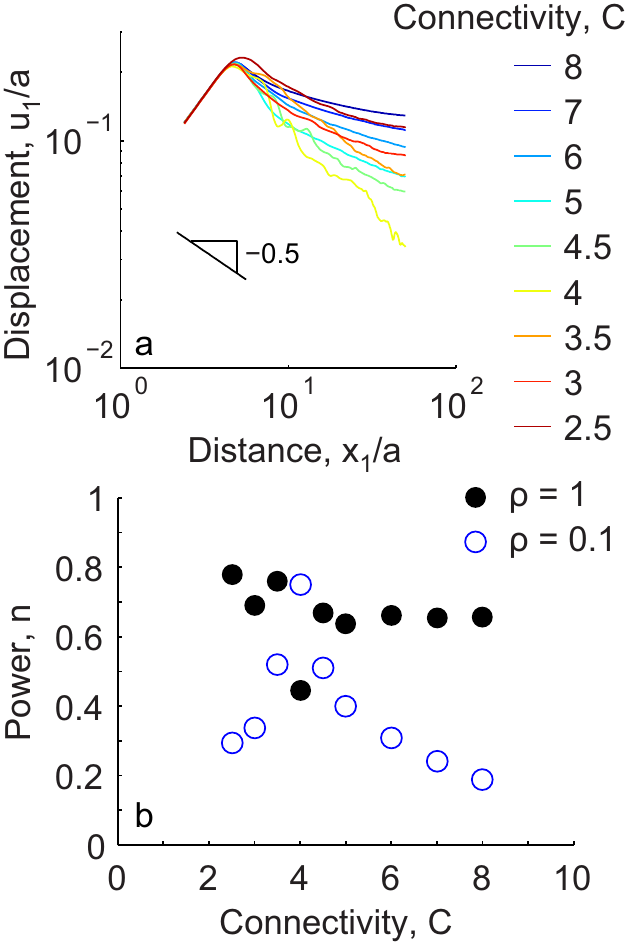}
\end{center}
\end{figure}
\noindent\textbf{Figure 4}: Simulated displacements due to an elongated cell. An ellipse with a ratio of semi-major to semi-minor axes $a_1/a_2=4$ is simulated contracting along its long axis in a circular region of radius $b=50a$ where $a\equiv\sqrt{a_1 a_2}$. (a) Displacements along the major axis $u_1$ are plotted against distance along the axis from the center of the ellipse $x_1$ for connectivities ranging from $C=2.5$ to $C=8$. As in Fig.~3, fluctuations are observed for connectivites near the critical value $C=4$. (b) Displacements far from the cell (i.e. for $x_1/a > 5$) are fit to $u_1=Ax_1^{-n}$. Solid black circles represent decay powers $n$ for fibers that do not buckle ($\rho=1$); open blue circles represent $n$ for fibers that do buckle ($\rho=0.1$). For simulations with buckling, except near the critical connectivity $C=4$, decay powers are smaller than the linear elastic solution $n=1$ and smaller than simulations without buckling.

\newpage
\begin{figure}[h!]
\begin{center}
\includegraphics[keepaspectratio=true,height=3.2 in]{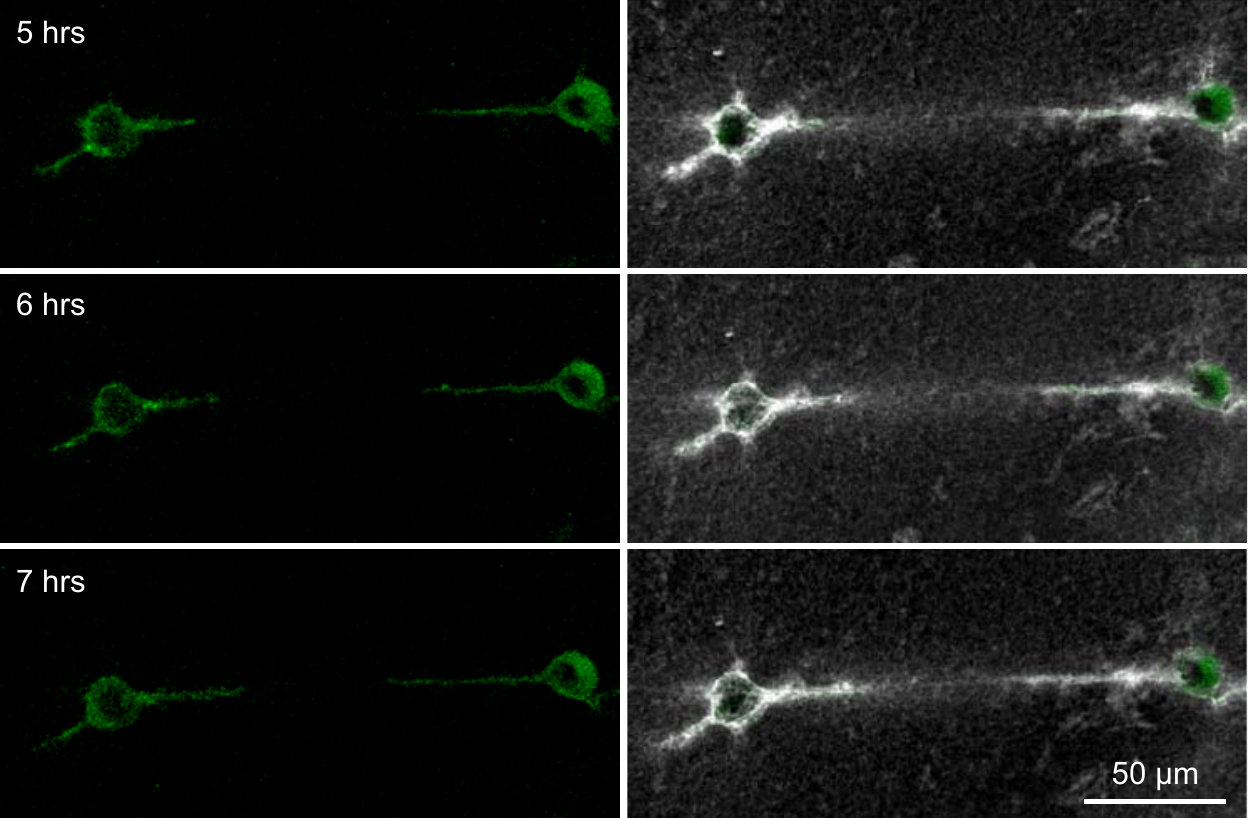}
\end{center}
\end{figure}
\noindent\textbf{Figure 5}: Pairs of cells spread toward one another along tethers. Panels in the left column show the cells (green), and panels in the right column show the cells (green) with the matrix (gray/white) at the same time point. The cells apply tensile force to the fibrin matrix resulting in matrix tethers connecting the cells (white). These tethers have a high density of matrix fibers, as apparent by the bright fluorescent signal in the space between the cells. The cells then spread along these tethers. Times are hours after the cells were seeded in the fibrin matrix.

\newpage
\begin{figure}[h!]
\begin{center}
\includegraphics[height=3.8in,keepaspectratio=true]{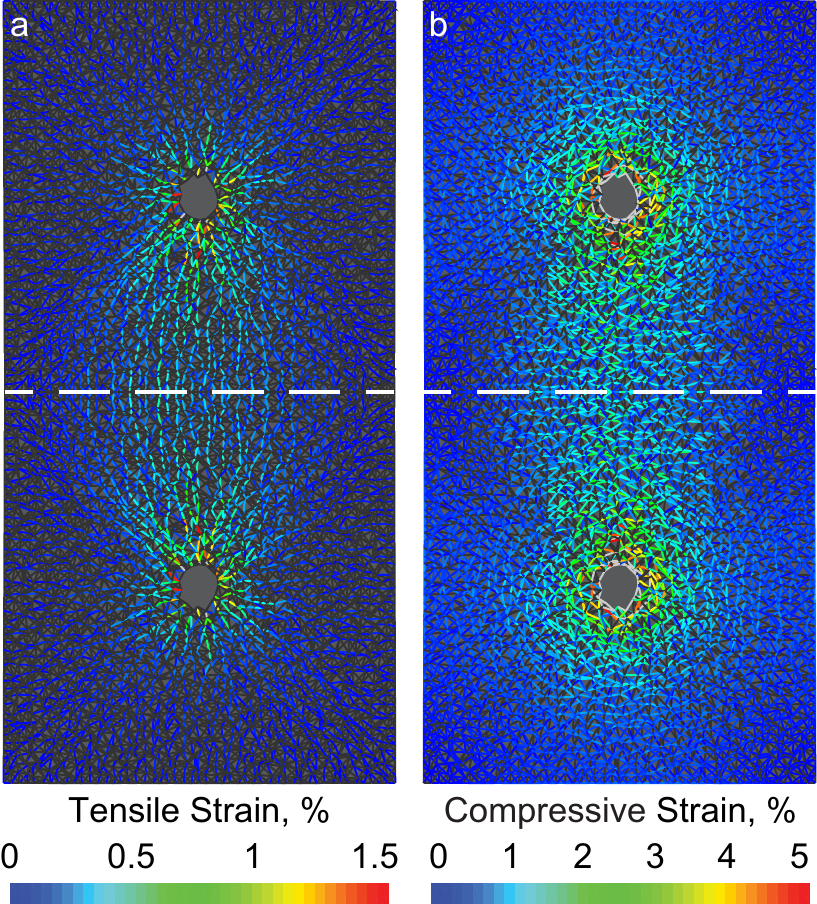}
\end{center}
\end{figure}
\noindent\textbf{Figure 6}: Fiber alignment and densification provide a mechanism for long-range cell mechanosensing. Mechanical interactions between cells are simulated using the FE model with a contracting circle and a symmetric boundary (dashed line). Plots show tensile (a) and compressive (b) strains within fibers. Fibers under tension (a) form intercellular tethers. Compressed fibers (b) are roughly perpendicular to tensile ones. The strains below the dashed line are the reflection of the strains above the dashed line.

\newpage

\section*{Supplemental Figures}

\vspace{11pt}

\begin{figure}[h!]
\begin{center}
\includegraphics[keepaspectratio=true,height=3.5 in]{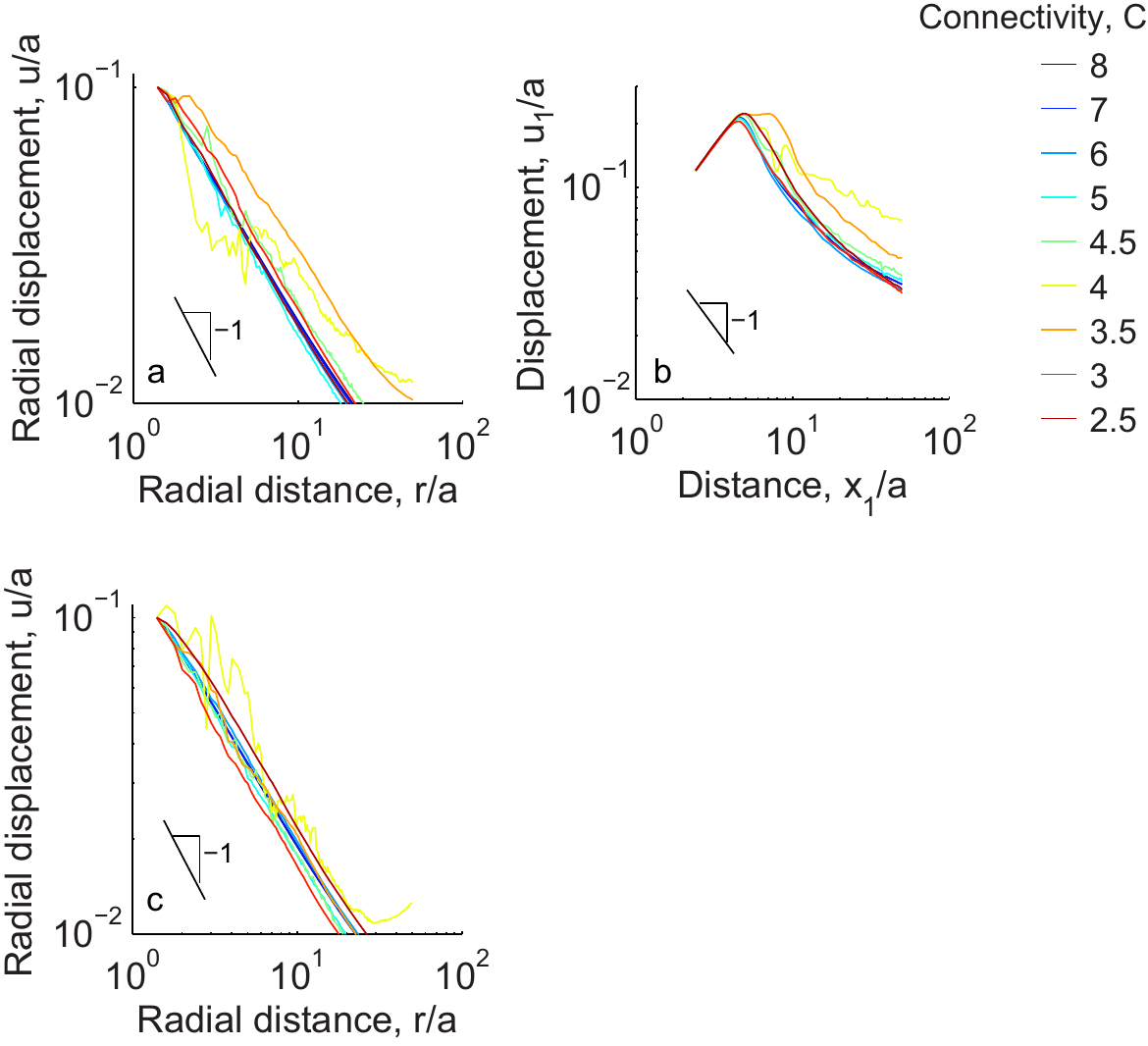}
\end{center}
\end{figure}
\noindent\textbf{Figure S1}: Simulated displacements for fibers that support compression, $\rho=1$. (a) Displacements due to a contracting circle are computed with the bilinear (Fig.~2a) model with $\rho=1$. The radial displacement component is averaged around circles of radius $r$ from the origin and plotted. Results show displacement $u$ vs. distance $r$ for simulations that used connectivities ranging from $C=2.5$ to $C=8$. See Fig. 3b for the case of microbuckling, $\rho=0.1$. (b) Displacements due to an ellipse with a ratio of semi-major to semi-minor axes $a_1/a_2=4$ with $\rho=1$. The displacements $u_1$ along the major axis are plotted against distance along the axis $x_1$ for connectivities ranging from $C=2.5$ to $C=8$. See Fig. 4a for the case of microbuckling, $\rho=0.1$. (c) Displacements due to a contracting circle computed with the strain stiffening model (Fig.~2b) with $\rho=1$. As in (a), the radial displacement component is averaged around circles of radius $r$ from the origin and plotted. See Fig.~3e for the case of microbuckling, $\rho=0.1$. For all cases, typical slopes are $-1$ on logarithmic axes, indicate displacements scale according to the 2D linear elastic solution, $u\!\sim\!1/r$.

\newpage

\begin{figure}[h!]
\begin{center}
\includegraphics[keepaspectratio=true,width=6.5 in]{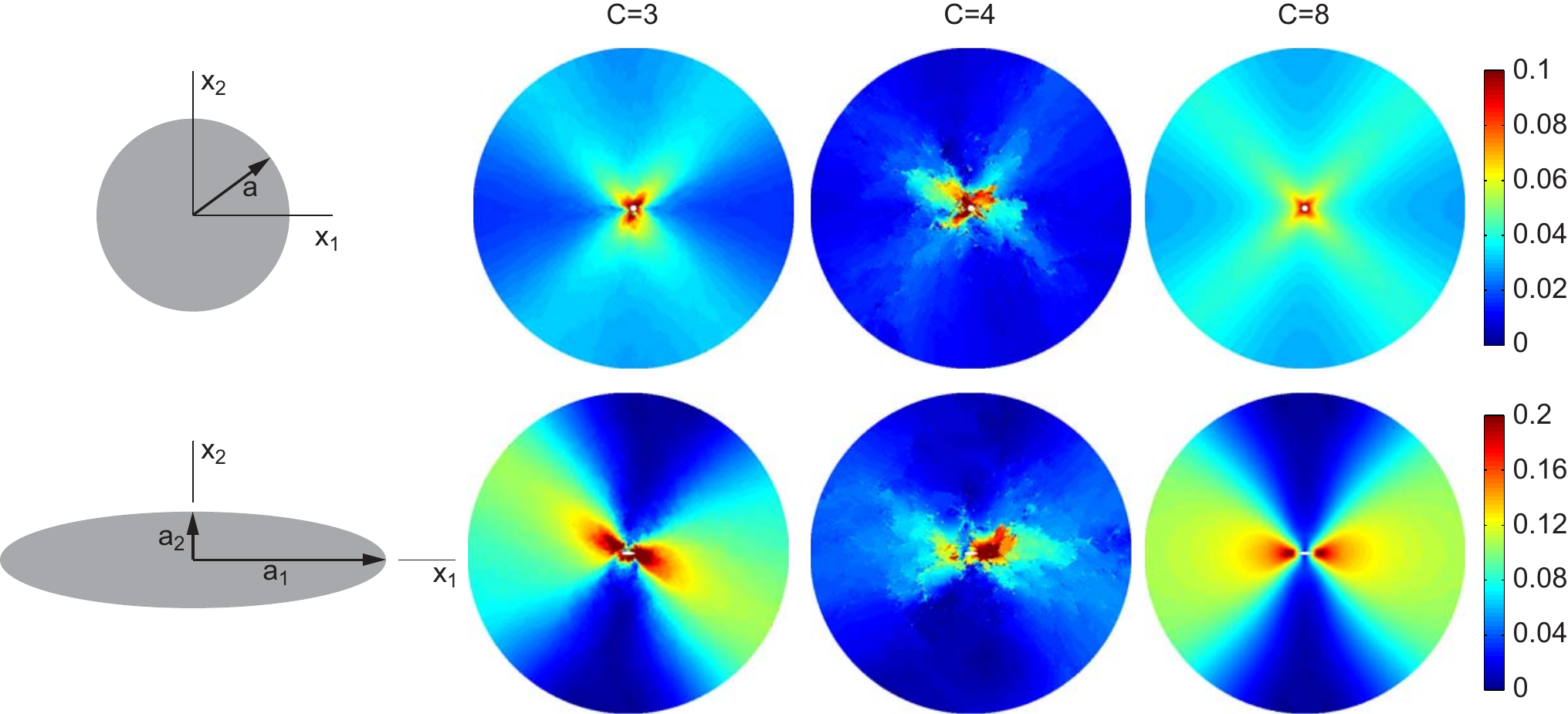}
\end{center}
\end{figure}
\noindent\textbf{Figure S2}: Displacements induced by a contracting circle and a contracting ellipse. A circle of radius $a$ (top row) or an ellipse with semi-major and semi-minor axes $a_1$ and $a_2$ (bottom row) is simulated in a circular region with radius $b=50a$. (For the ellipse, $a \equiv \sqrt{a_1 a_2}$.) For the circle, contractile displacements are applied uniformly around the perimeter, $u(r)=-0.1a$ at $r=a$ where $r=\left(x_1^2 + x_2^2\right)^{1/2}$. For the ellipse, contractile displacements are applied only along the major axis, $u_1(x_1) = -0.1a(x_1/a_1)$. The outer boundary $r=b$ is free of applied tractions. Microbuckling is simulated using the bilinear model with $\rho=0.1$. Displacement magnitudes normalized by $a$ are shown for the contracting circle and ellipse for connectivities $C$ of 3, 4, and 8. Near the critical connectivity $C=4$ large fluctuations in displacements occur for both the contracting circle and the ellipse, in agreement with previous models \cite{wyart,broedersz2011}. These fluctuations are not present at lower ($C=3$) or higher ($C=8$) connectivities, where displacement fields are smoother.

\newpage

\begin{figure}[h!]
\begin{center}
\includegraphics[keepaspectratio=true,height=1.7 in]{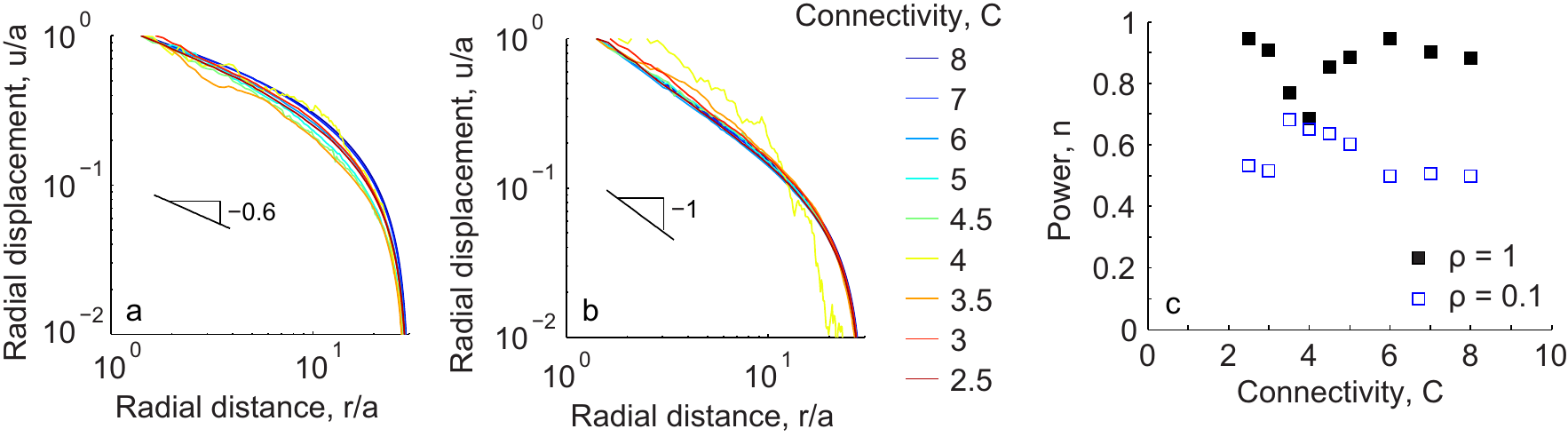}
\end{center}
\end{figure}
\noindent\textbf{Figure S3}: Effect of boundary conditions on simulation results. The simulations of Fig. 3 are repeated with fixed boundaries instead of free. (a) For elements with microbuckling ($\rho=0.1$), displacements for all connectivities $C$ have a slope of $\sim\!-0.6$ on logarithmic axes. (b) The simulations are repeated for elements without microbuckling ($\rho=1$), and displacements have a slope of $\sim\!-1$ on a logarithmic scale. (c) For each connectivity, and for simulations with microbuckling ($\rho=0.1$) and without ($\rho=1$), displacements are fit to the linear elastic solution for a circular region of finite radius, $u=Ar^{-n}+Br^n$. (No fitting is performed to $u=Ar^{-n}$, because, as shown in (a) and (b), the fixed boundary affects the propagation of displacements for $r/a>10$.) The fit power $n$ is plotted for all cases. Similar to simulations with free boundaries (Fig. 3), simulations with microbuckling ($\rho=0.1$) have lower powers of $n$, indicating displacements propagate over a long range when microbuckling is present.

\newpage

\begin{figure}[h!]
\begin{center}
\includegraphics[keepaspectratio=true,height=2.0 in]{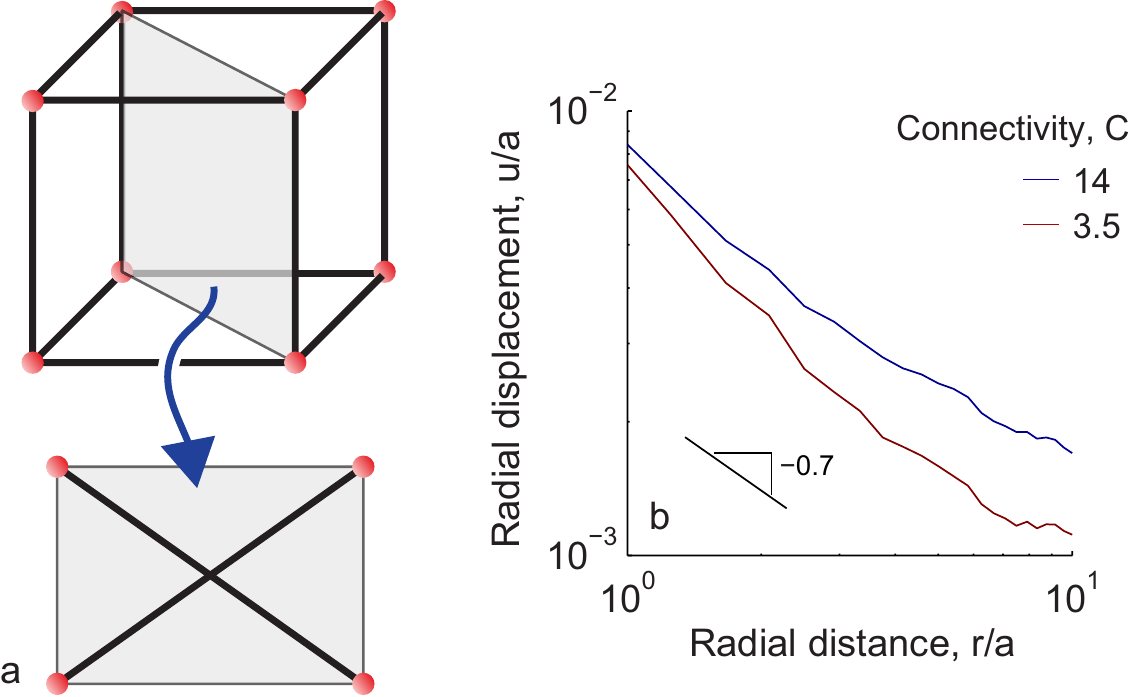}
\end{center}
\end{figure}
\noindent\textbf{Figure S4}: The displacements due to a contracting sphere in a fibrous matrix are simulated using a 3D model. $49\!\times\!49\!\times\!49$ nodes are used in a $20a\times20a\times20a$ region, where $a$ is the radius of the sphere. A symmetric boundary is used at the bottom of the cubic region ($z = -10a$), and other boundaries are free. An inward displacement of $0.1a$ is applied to the nodes located at $r=a$. (a) Fiber connectivity. Each cube of $8$ nodes is connected along the the cube's edges. Additionally, elements connect the diagonals as shown in the sketch. To simplify visualization, the sketch shows connections between only $2$ diagonals, but the model connects all $4$ diagonals with elements. As with the 2D models, lower connectivity is simulated by randomly selecting elements to delete. Deleted elements are  replaced by weak elements with stiffness six orders of magnitude lower than that of the deleted elements. (b) Displacements due to the contracting sphere are averaged along circles of radius $r$ from the center of the sphere in the $x\!-\!y$ plane and plotted against radial distance for connectivities $C$ of $3.5$ (below the critical value of $6$) and $14$ (full connectivity). Fits to $u=Ar^{-n}$ give $n=0.82$ and $0.67$ for $C=3.5$ and $14$, respectively.

\newpage

\begin{figure}[h!]
\begin{center}
\includegraphics[keepaspectratio=true,height=3.8 in]{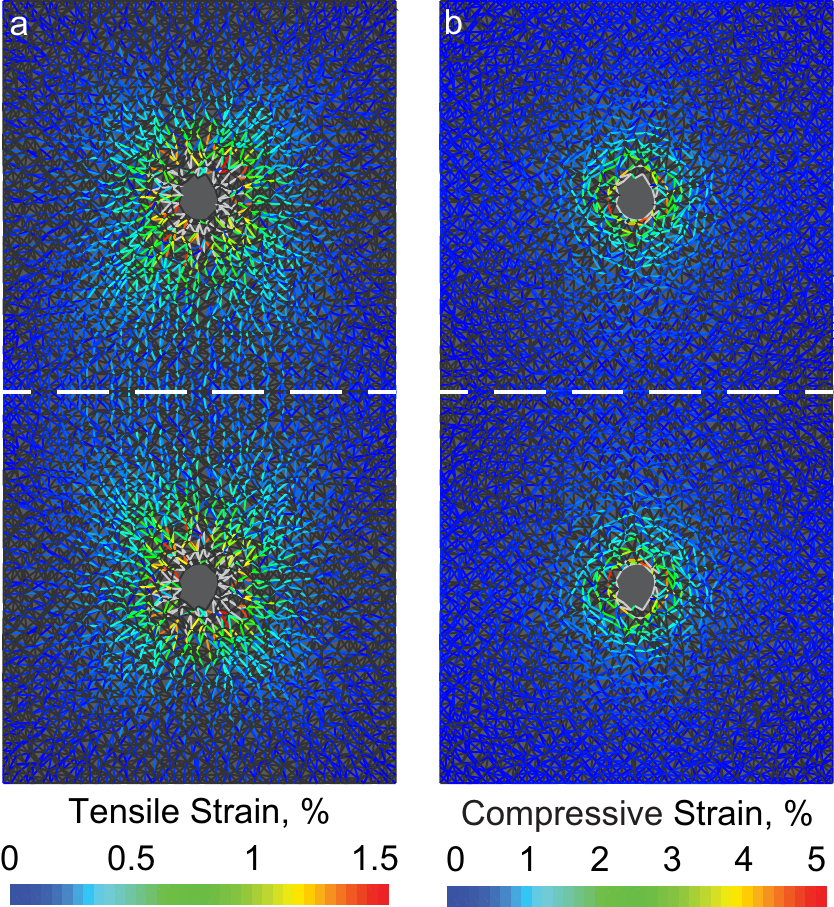}
\end{center}
\end{figure}
\noindent\textbf{Figure S5}: Tethers do not form when the matrix resists compression. Plots show tensile (a) and compressive (b) strains for the same simulation as in Fig. 5, but for fibers with equal stiffness in compression and tension ($\rho=1$). Tensile strains (a) propagate radially outward from the contracting circle with no preferred directionality, and therefore no tethers form. Compressive strains (b) are roughly perpendicular to tensile strains.

\newpage

\begin{figure}[h!]
\begin{center}
\includegraphics[keepaspectratio=true,height=1.7 in]{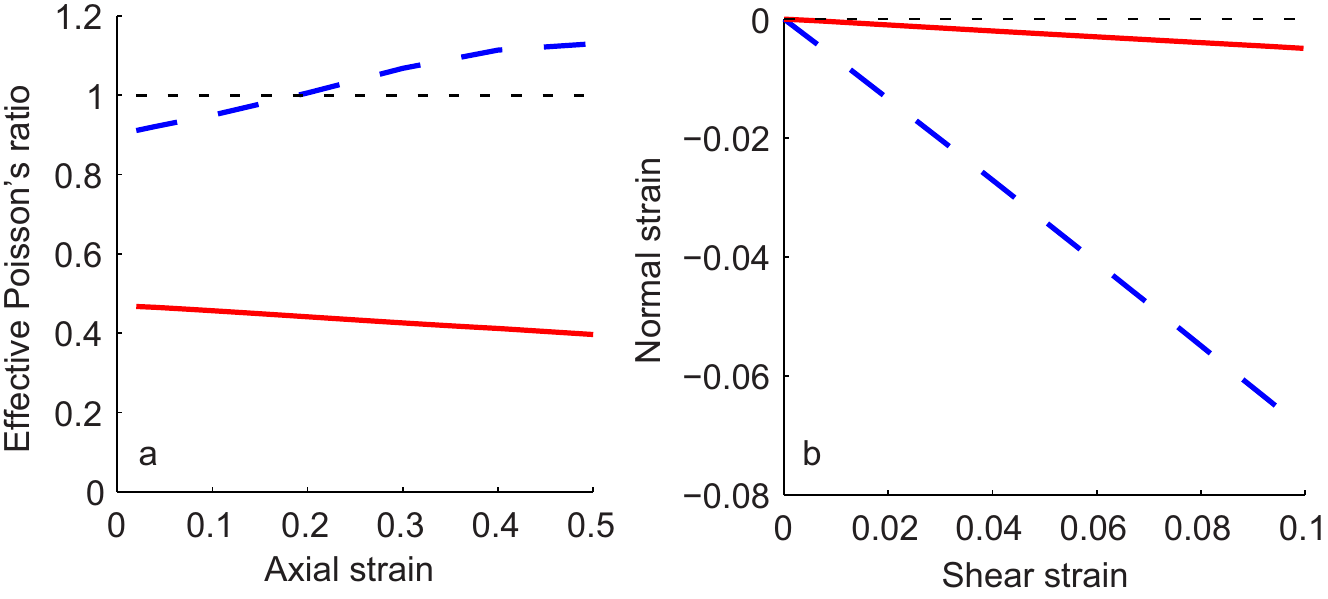}
\end{center}
\end{figure}
\noindent\textbf{Figure S6}: Microbuckling induces negative compressibility in tension and negative normal strains in shear. (a) The effective (engineering) Poisson's ratio is calculated in a uniaxial tension simulation for a model with matrix fibers that support compression ($\rho=1$, solid red line) or buckle ($\rho=0.1$, dashed blue line). In two dimensions, Poisson's ratios greater than 1 (dashed black line) indicate negative compressibility, in agreement with an experimental study on fibrin \cite{12brown}.
(b) Negative normal strains are observed in a simulation under shear loading for matrix fibers that support compression ($\rho=1$, solid red line) or buckle ($\rho=0.1$ dashed blue line). Similar to \cite{11janmey},
negative normal strains are significantly larger for elements that simulate buckling. Simulations in this figure use the bilinear stress--strain relationship, but results are nearly identical for the WLC relationship. Results shown are for connectivity of $C=8$.

\newpage

\begin{figure}[h!]
\begin{center}
\includegraphics[keepaspectratio=true,height=1.7 in]{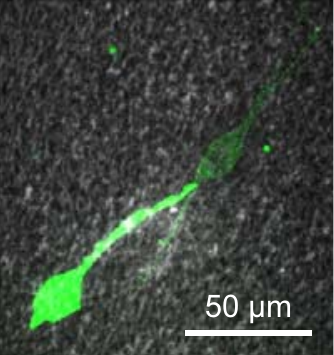}
\end{center}
\end{figure}
\noindent\textbf{Figure S7}: Pairs of fibroblast cells (green) in a 3D fibrin matrix (white) occasionally spread until they touch their neighbors. Image was captured with confocal microscopy 14 hrs after seeding the cells in the fibrin matrix.

\newpage

\section*{Supplemental Data}

\noindent Experimental data of 3D cell-induced displacements (Fig.~1b) for $n=6$ cells are below. The data report the magnitude of cell-induced matrix displacements after averaging along different paths outward from the cell over time. For more details on how these cell-induced displacements are measured, see the methods section. The data set for each cell is displayed as an array of ordered pairs, ($r$,$u$) where each pair gives the distance from the cell's center $r$ and the magnitude of the cell-induced matrix displacement $u$.

\subsection*{Cell 1.} (25,0.68) (29,0.64) (33,0.61) (38,0.57) (42,0.53) (46,0.50) (50,0.47) (55,0.43) (59,0.40) (63,0.37) (67,0.35) (72,0.35) (76,0.38) (80,0.35) 

\subsection*{Cell 2.} (15,1.91) (20,1.78) (24,1.60) (28,1.42) (32,1.25) (36,1.15) (40,1.08) (44,1.01) (48,0.97) (52,0.90) (56,0.81) (60,0.69) (64,0.58) (68,0.53) (72,0.54) 

\subsection*{Cell 3.} (16,1.14) (21,1.04) (26,0.95) (31,0.89) (35,0.83) (40,0.77) (45,0.73) (49,0.70) (54,0.67) (59,0.66) (64,0.65) (68,0.64) (73,0.65) 

\subsection*{Cell 4.} (17,1.33) (22,1.22) (26,1.14) (31,1.07) (36,1.00) (41,0.95) (46,0.91) (51,0.87) (55,0.84)
 
\subsection*{Cell 5.} (14,1.89) (18,1.60) (21,1.40) (25,1.30) (29,1.25) (32,1.21) (36,1.19) (40,1.17) (43,1.16) (47,1.14) (51,1.12) 

\subsection*{Cell 6.} (12,1.76) (16,1.66) (20,1.34) (24,1.25) (27,1.07) (31,1.18) (35,1.12) (39,1.08) (43,1.04) (46,0.99) (50,0.95) (54,0.92) (58,0.89) (61,0.86) (65,0.83)

\end{document}